\documentclass[USenglish,cleveref,autoref,thm-restate]{lipics-v2021} 
\nolinenumbers 
\pdfoutput=1 
\usepackage{mathtools}
\usepackage{cite}
\usepackage{xparse}
\usepackage{algorithm}
\usepackage{algpseudocode}
\usepackage{thm-restate}
\newcommand{\ang}[1]{\langle #1\rangle}

\renewcommand{\ang}[1]{\langle #1\rangle}
\newcommand{\RE}{\mathbb{R}}            
\newcommand{\eps}{\varepsilon}          
\newcommand{\ST}{\,:\,}                 
\newcommand{\bd}{\partial\kern+1pt}     

\newcommand{\SP}{\kern+1pt}             
\newcommand{\NP}{\kern-1pt}             
\newcommand{\PERP}{\kern-2pt \perp \kern-2pt} 

\newcommand{\vctr}[1]{\mathbf{#1}}      
\newcommand{\mat}[1]{\mathbf{#1}}       
\newcommand{\TT}{\top}                  
\DeclareMathOperator{\interior}{int}    
\DeclareMathOperator{\closure}{cl}      
\DeclareMathOperator{\cp}{cp}           

\newcommand{\hilb}{d^H_\Omega}
\newcommand{\hball}{B^H_\Omega}
\newcommand{\hyp}[1]{{{#1}}}
\newcommand{\norm}[1]{\lVert {#1}\rVert}
\NewDocumentCommand{\distH}{O{}}{d^{H}_{#1}} 
\NewDocumentCommand{\distF}{O{}}{d^{F}_{#1}}
\NewDocumentCommand{\distRF}{O{}}{d^{\kern+1pt r \kern-1pt F}_{#1}}
\NewDocumentCommand{\distG}{O{}}{d^{*}_{#1}} 
\NewDocumentCommand{\ballH}{O{}}{B^{H}_{#1}} 
\NewDocumentCommand{\ballF}{O{}}{B^{F}_{#1}}
\NewDocumentCommand{\ballRF}{O{}}{B^{rF}_{#1}}
\NewDocumentCommand{\ballG}{O{}}{B^{*}_{#1}} 
\NewDocumentCommand{\closestH}{O{}}{\cp^{H}_{#1}} 
\NewDocumentCommand{\closestF}{O{}}{\cp^{F}_{#1}}
\NewDocumentCommand{\closestRF}{O{}}{\cp^{rF}_{#1}}

\usepackage{xcolor}

\title{Classifiers in High Dimensional Hilbert Metrics}
\titlerunning{Classifiers in High Dimensional Hilbert Metrics}

\author{Aditya Acharya}{Department of Computer Science, University of Maryland, College Park, USA \and \url{https://www.cs.umd.edu/~acharya/}}{adach@umd.edu}{https://orcid.org/0000-0002-0359-1913}{}

\author{Auguste H. Gezalyan}{Department of Computer Science, University of Maryland, College Park, USA \and \url{~}}{octavo@umd.edu}{https://orcid.org/0000-0002-5704-312X}{}

\author{David M. Mount}{Department of Computer Science, University of Maryland, College Park, USA \and \url{https://www.cs.umd.edu/~mount/}}{mount@umd.edu}{https://orcid.org/0000-0002-3290-8932}{}

\authorrunning{A. Acharya, A. H. Gezalyan, and D. M. Mount}

\Copyright{Aditya Acharya, Auguste H. Gezalyan, and David M. Mount}


\begin{document}

\maketitle

\begin{abstract}
Classifying points in high dimensional spaces is a fundamental geometric problem in machine learning. In this paper, we address classifying points in the $d$-dimensional Hilbert polygonal metric. The Hilbert metric is a generalization of the Cayley-Klein hyperbolic distance to arbitrary convex bodies and has a diverse range of applications in machine learning and convex geometry. We first present an efficient LP-based algorithm in the metric for the large-margin SVM problem. Our algorithm runs in time polynomial to the number of points, bounding facets, and dimension. This is a significant improvement on previous works, which either provide no theoretical guarantees on running time, or suffer from exponential runtime. We also consider the closely related Funk metric. We also present efficient algorithms for the soft-margin SVM problem and for nearest neighbor-based classification in the Hilbert metric.
\end{abstract}

\section{Introduction} \label{sec:intro}

In machine learning, \emph{binary classification} is a supervised learning task where an algorithm learns to map input data (features points) to one of two mutually exclusive classes or labels, often denoted as \emph{positive} and \emph{negative}. The primary objective is to train a model that can accurately predict which of the two predefined classes a new, unseen data point belongs to. Binary classification is one of the most fundamental geometric problems in machine learning~\cite{cristianini2000introduction, steinwart2008support, russell2016artificial, bishop2016pattrec} as it is related to a wide range of empirical decision problems. 

Support vector machines (SVMs) were introduced by Vapnik and Chervonenkis~\cite{vapnik1964svm} as a type of binary classification algorithm~\cite{cortes1995support}. We are given two sets of \emph{training points} $P^+$ and $P^-$, representing feature vectors in $\RE^d$, that are sampled from two distinct spatial distributions. SVM solves this classification problem by computing a hyperplane that separates the two training sets, such that the minimum distance from each point set to the separating hyperplane, called the \emph{margin}, is as large as possible. 

SVMs are widely used in a diverse set of fields such as computer vision~\cite{szeliski2022computer, krig2014computer}, natural language processing~\cite{decoste2002training, joachimslearning}, and computational biology~\cite{huang2018applications, ben2008support}. They have been well studied in Euclidean and hyperbolic geometry~\cite{cho2019large}, but classification in other non-Euclidean geometries is less well studied. In this paper, we study this problem from the perspective of the Hilbert geometry and the closely related Funk geometry.

The Hilbert metric (which will be defined in Section~\ref{sec:prelim}) is a generalization of the Cayley-Klein model of hyperbolic geometry to arbitrary convex sets. It has several desirable properties, such as having straight line geodesics and being invariant under projective transformations \cite{papadopoulos2014handbook}. The Hilbert metric has a diverse array of applications including: quantum information theory \cite{reeb2011hilbertquantum}, real analysis \cite{birkhoff1957extensions}, and machine learning \cite{nielsen2022nonlinear, nielsen2019clustering, karwowski2025hilbert}. Note that the hyperbolic metric (under the Beltrami-Klein model) is a special case of the Hilbert metric when the defining convex body is a Euclidean ball.

The Hilbert and Funk geometries are applicable whenever the domain of interest is a convex body and distances are sensitive to the proximity to the boundary. As an example, consider objects that are discrete probability distributions over a domain of size $m$. Each point in this space is a vector $(p_1, \ldots, p_m)$, where $0 \leq p_i \leq 1$ and $\sum_{i=1}^m p_i = 1$. These constraints imply that the points lie within a convex domain, specifically the $(m-1)$-dimensional probability simplex. Euclidean distances are not well suited for this context, because they are not sensitive to variations in probabilities that are very close to the critical values of $0$ and $1$ as compared with those that are far from these extremes. In contrast, the Hilbert distance is very sensitive as points approach the boundary. Nielsen and Sun provided empirical evidence that the Hilbert metric is competitive with other widely-used distance functions for clustering in the probability simplex~\cite{nielsen2019clustering}. Other applications of Hilbert and related metrics include the analysis of networks through hyperbolic geometry~\cite{KPK10}, the analysis of positive definite matrices in deep learning \cite{ISY11, LDL23}, and lattice-based cryptosystems~\cite{AFM24}.

\subsection{Related Work}

In the Euclidean metric, SVMs are typically solved by formulating them as a convex optimization problem \cite{SVM-convex-opt}. The authors of \cite{cho2019large} were among the first to consider SVMs in hyperbolic geometry. They demonstrated that, in practice, their approach significantly outperforms Euclidean classifiers on various benchmarks by respecting the intrinsic geometry of the data. However, their method involved solving a non-convex optimization problem, unlike the convex formulation of Euclidean SVMs. Hence, without careful initialization, it neither guarantees a global optimum nor a worst-case polynomial running time. 

Recently, the authors of \cite{AGV25} presented an algorithm for the same polytopal SVM problem that we consider here. They showed that SVM could be modeled as an LP-type problem. Their algorithm's running time is linear in the number of points but grows polynomially in the total complexity of the polytope defining the metric. Unfortunately, the total complexity of this polytope generally grows exponentially in the dimension of the space. As such, their method remains unsuitable for high dimensions, where most applications in machine learning lie. In contrast, in this work, our running times are polynomial in dimension. 

\subsection{Our Contributions}

In this paper, we explore an alternative viewpoint of Hilbert geometry to develop an efficient algorithm for the Hilbert SVM problem. The main feature of our approach is that the running time increases only polynomially with the dimension of the space. Further, our algorithms involve reductions to linear programming, for which efficient and practical software solutions exist. Because of our reliance on numerical solutions to linear programming, we need to assume that the number of bits needed to represent our inputs is bounded, and our output is computed to within a given user-supplied approximation error. The following theorem formalizes our main result.

\begin{theorem}\label{thm:main-result}
Consider a polytope $\Omega$ in $\RE^d$ defined by $m$ hyperplanes, and two linearly separable point sets $P^+$ and $P^-$ of total size $n$ contained within $\Omega$'s interior. Assume further that all of the defining coordinates and coefficients are representable as $B$-bit rational numbers. Then there exists an algorithm running in time $O(\text{poly}\,(n,m,d,B,\log(1/\eps))$ that computes a separating hyperplane whose margin in the Hilbert metric defined by $\Omega$ is within an additive error $\eps$ of the optimum.
\end{theorem}

The notable feature of our approach is the polynomial dependence on the dimension of the space, which is a crucial assumption in the applications we have in mind. We also generalize our results to the closely related Funk geometry (defined in Section~\ref{sec:hilbert-def}). 

In addition to the linearly separable case, we also provide a \emph{soft-margin} formulation for the SVM problem, which applies when the two point set are not linearly separable (Section \ref{sec:soft-margin}). Finally, we conclude with a nearest neighbor-based classifier when the Hilbert metric is embedded in a normed space (Section \ref{sec:nn-class}). Note that, in contrast with the Euclidean case where bisectors are hyperplanes, Hilbert bisectors are not hyperplanes. Therefore, this problem cannot be reduced to the SVM problem.

\section{Mathematical Preliminaries} \label{sec:prelim}

In this section, we provide a number of definitions and facts, which will be used throughout. We will use different fonts, specifically $p$ and $\vctr p$, to distinguish between a point $p \in \RE^d$ and its associated $d$-dimensional column vector $\vctr p$. We represent a $(d-1)$-dimensional flat, that is a hyperplane, $\hyp L$ in $\RE^d$ as a pair $(\vctr w, c) \in \RE^d \times \RE$ represented by the function $f_{\hyp L}(x) = \vctr w^{\TT} \vctr{x} + c$. A hyperplane partitions space into three sets, $\hyp L^-$, $\hyp L$, $\hyp L^+$, where a point $\vctr x \in \RE^d$ belongs to each of these sets depending on whether $f_L(x)$ is $< 0$, $= 0$, or $> 0$, respectively.

Given a set of $m$ hyperplanes, $\mathcal{L} = \{\hyp L_i : (\vctr w_i, c_i)\}_{i=1}^m$ let $\Omega = \Omega(\mathcal{L})$ denote the closure of the space bounded by the positive half-spaces $\hyp L_i^+$, that is, $\Omega = \closure\!\big(\bigcap_{i=1}^{m} \hyp L_i^+\big)$. Let $\bd \Omega$ denote the boundary of $\Omega$. Throughout, we assume that $\Omega$ is bounded and full dimensional. Given a nonnegative integer $m$, let $[m]$ denote the index set $\{1, \ldots, m\}$.

Given two points $p, q \in \interior(\Omega)$, let $\overline{p q}$ denote the \emph{chord} of $\Omega$ passing through these points, that is, the intersection of $\Omega$ with the line through $p$ and $q$. Let $\overleftrightarrow{p q}$ denote the line that passes through them, and $\overrightarrow{pq}$ denote the ray starting from $p$ and passing through $q$.

Given $p, q \in \RE^d$, let $d(p,q)$ denote their Euclidean distance. Given a hyperplane $\hyp L$, let $d(p, \hyp L)$ denote the minimum Euclidean (perpendicular) distance from $p$ to $\hyp L$.

\subsection{Hilbert Geometry} \label{sec:hilbert-def}

\begin{figure}[htbp]
  \centerline{\includegraphics[scale=0.40,page=1]{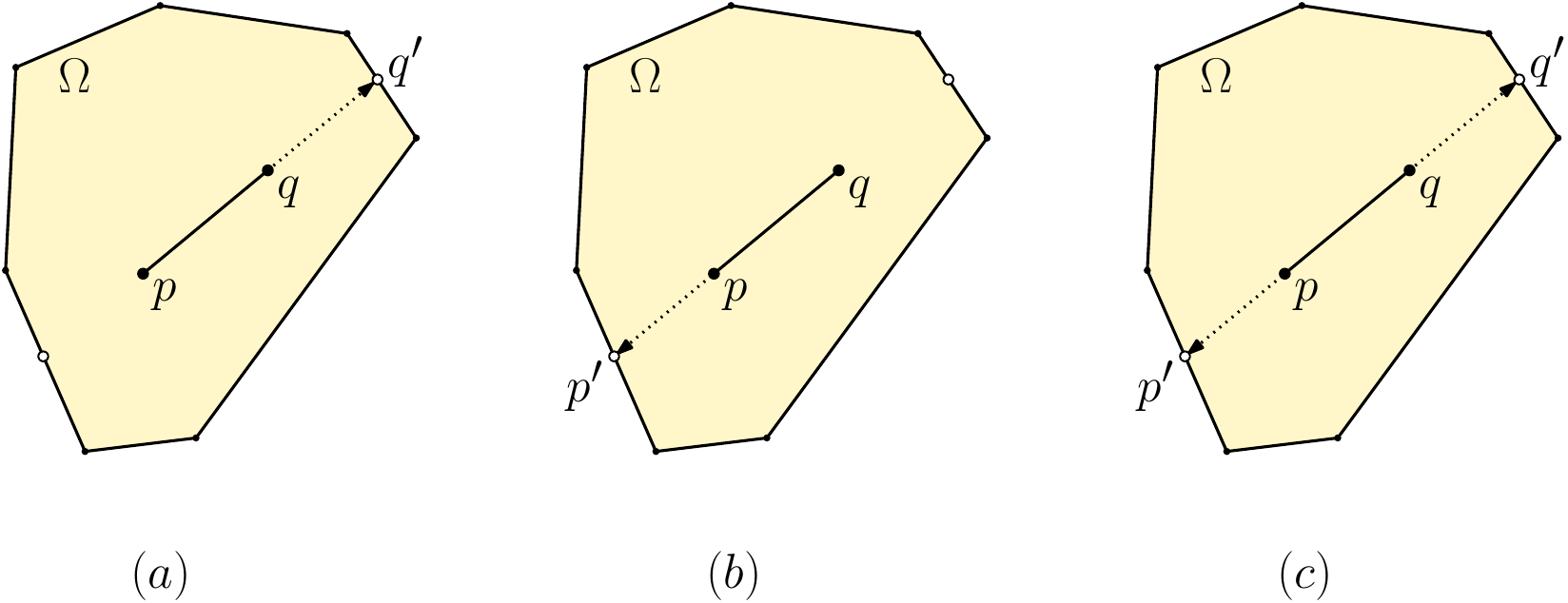}}
  \caption{The (a) forward Funk, (b) reverse Funk, (c) and Hilbert distances between $p$ and $q$ in $\Omega$.}
  \label{fig:funk-hilbert-1}
\end{figure}

In this section we introduce the Funk and Hilbert distances. Given a convex body $\Omega$ and any two points $p, q \in \interior(\Omega)$, if $p = q$ all the distances are defined to be zero. Otherwise, let $p'$ and $q'$ denote the endpoints of the chord $\overline{p q}$ ordered as $\ang{p', p, q, q'}$ (see Figure~\ref{fig:funk-hilbert-1}). The Funk and Hilbert distances are defined:
\begin{eqnarray}
    \text{Funk:}~~~
        \distF[\Omega](p,q)
            & := & \ln \frac{\|p - q'\|}{\|q - q'\|} \label{eq:funk} \\
    \text{Reverse Funk:}~~~
        \distRF[\Omega](p,q)
            & := & \distF[\Omega](q,p)
            ~ = ~ \ln \frac{\|q - p'\|}{\|p - p'\|} \label{eq:rev-funk} \\[5pt]
    \text{Hilbert:}~~~
        \distH[\Omega](p,q)
            & := & \frac{\distF[\Omega](p,q) + \distRF[\Omega](p,q)}{2}
            ~ = ~ \frac{1}{2} \ln \frac{\|q - p'\|}{\|p - p'\|} \frac{\|p - q'\|}{\|q - q'\|}. \label{eq:hilbert}
\end{eqnarray}
Intuitively, $\distF[\Omega](p,q)$ is a measure of how far $q$ is from $p$ relative to the boundary that lies beyond $q$, and $\distRF[\Omega](p,q)$ is just the reverse of this. The Hilbert distance $\distH[\Omega](p,q)$ symmetrizes these by taking their arithmetic mean.

These distance functions are all nonnegative and satisfy the triangle inequality~\cite{papadopoulos2014handbook}. The Hilbert distance defines a \emph{metric} over $\interior(\Omega)$. The Funk and reverse Funk are asymmetric and hence define \emph{weak metrics} over $\interior(\Omega)$. Observe that as either point approaches the boundary of $\Omega$, the Hilbert distance approaches $\infty$. The Hilbert distance is invariant under invertible projective transformations.

\subsection{Birkhoff's Formulation and Metric Balls}

Birkhoff~\cite{birkhoff1957extensions} provided an alternative, equivalent definition of these metrics, which is more suited to our approach.

\begin{proposition}[Birkhoff's Formulation] \label{prop:birkhoff}
Given a set of $m$ hyperplanes, $\mathcal L = \{\hyp L_i : (\vctr w_i, c_i)\}_{i=1}^m$, then for any pair of points $p,q \in \interior(\Omega(\mathcal L)$:
\begin{align*}
    \distF[\Omega](p,q) 
        & ~ = ~ \max_{i\in[m]} \log \frac{d(p,L_i)}{d(q,L_i)} 
          ~ = ~ \max_{i\in[m]} \log \frac{\vctr w_i^\TT \vctr p + c_i}{\vctr w_i^\TT \vctr q + c_i} \\
    \distRF[\Omega](p,q) 
        & ~ = ~ \max_{i\in[m]}\log \frac{d(q,L_i)}{d(p,L_i)} 
          ~ = ~ \max_{i\in[m]}\log \frac{\vctr w_i^\TT \vctr q + c_i}{\vctr w_i^\TT \vctr p + c_i} \\
    \hilb(p,q) 
        & ~ = ~ \frac{1}{2} \max_{j,k\in[m]} \log \frac{d(p,L_j)}{d(q,L_j)} \frac{d(q,L_k)}{d(p,L_k)} 
          ~ = ~ \frac{1}{2} \max_{j,k\in[m]} \log \left(\frac{\vctr w_j^\TT \vctr p + c_j}{\vctr w_j^\TT \vctr q + c_j} \frac{\vctr w_k^\TT \vctr q + c_k}{\vctr w_k^\TT \vctr p + c_k}\right).
\end{align*}
\end{proposition}

This follows from the fact that $\Omega$ is convex, and observing that if $\overrightarrow {pq}$ intersects $\bd\Omega$ in $q'$, on the bounding hyperplane $L_k$, then by similar triangles 
\[
    \frac{d(p,q')}{d(q,q')}
        ~ = ~ \frac{d(p,L_is)}{d(q,L_i)}.
\]
Intuitively, this provides a way to define the metrics without explicitly calculating where the line $\overleftrightarrow{pq}$ intersects $\bd \Omega$. One consequence of the proposition is that the metrics can be reduced to a sup norm over $\RE^t$ for some finite dimension $t$, using a non-linear mapping function \cite{supnorm_m2}. The other useful consequence is its application in constructing metric balls in high dimensions.

Letting $\distG[\Omega]$ denote any of the above metrics and given a point $p \in \interior(\Omega)$ and scalar $r \geq 0$, the associated metric balls are defined in the standard way.
\[
    \ballG[\Omega](p,r) 
        ~ := ~ \{ q \in \Omega \ST \distG[\Omega](p,q) \leq r \}.
\]
It is well known that the Funk metric about a point $p$ is a scaling of $\Omega$ about $p$ by an appropriate factor that depends on the radius. This can be derived directly from Proposition~\ref{prop:birkhoff}:
\begin{align}
    \ballF[\Omega](p,r)
        & ~ = ~ \left\{ q \in \Omega \ST \distF[\Omega](p,q) \leq r \right\} \notag\\
        & ~ = ~ \left\{ q \in \Omega \ST \max_{i\in[m]}\log \frac{\vctr w_i^\TT \vctr p + c_i}{\vctr w_i^\TT \vctr q + c_i} \leq r \right\} \notag\\
        & ~ = ~ \left\{ q \in \Omega \ST \log \frac{\vctr w_i^\TT \vctr p + c_i}{\vctr w_i^\TT \vctr q + c_i} \leq r, \; \forall i\in[m]\right\} \notag\\
        & ~ = ~ \left\{ q \in \Omega \ST 0 \leq \vctr w_i^\TT \vctr q + c_i - e^{-r}(\vctr w_i^\TT \vctr p + c_i),\; \forall i\in[m] \right\} \notag\\
        & ~ = ~ \bigcap_{i\in[m]} \left\{ q \in \Omega \ST 0 \leq \vctr w_i^\TT \vctr q + c_i - e^{-r}(\vctr w_i^\TT \vctr p + c_i)\right\}. \label{eq:funk-ball-characterization}
\end{align}

\begin{figure}[htbp]
  \centerline{\includegraphics[scale=0.40,page=2]{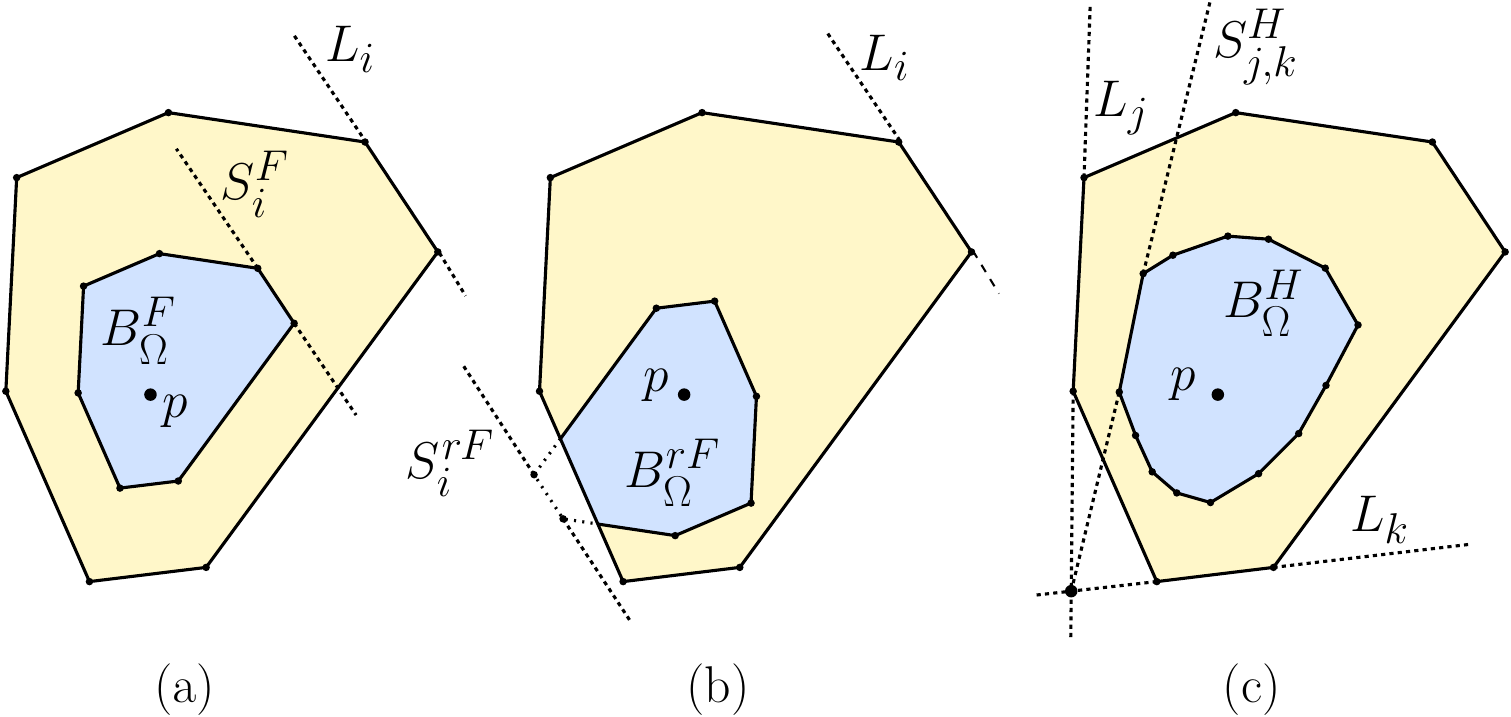}}
  \caption{Defining balls and their bounding hyperplanes for (a) Funk, (b) Reverse Funk, and (c) Hilbert.}
  \label{fig:funk-hilbert-2}
\end{figure}

Fixing $p$ and $r$, let $c_i^* := c_i - e^{-r}(\vctr w_i^\TT \vctr p + c_i)$, and let $S_i^F$ be the hyperplane represented by $(\vctr w_i, c_i^*)$ (See Figure~\ref{fig:funk-hilbert-2}(a)). Observe that $S_i^F$ is parallel to $L_i$. We have the following:

\begin{lemma}[Funk Balls] \label{lem:fball}
Given a set of $m$ hyperplanes, $\mathcal L$ as in Proposition~\ref{prop:birkhoff}, the Funk ball $\ballF[\Omega(\mathcal L)](p,r)$ is a polytope bounded by the $m$ hyperplanes, $S_i^F \ST (\vctr w_i, c_i^*)$, for $i\in[m]$.
\end{lemma}

A similar result can be proved for reverse-Funk balls (see Figure~\ref{fig:funk-hilbert-2}(b)). 

We next use Proposition~\ref{prop:birkhoff} to demonstrate the Hilbert balls are convex polytopes, but the number of bounding facets is larger by the square. Following the same reasoning as in Eq.~\eqref{eq:funk-ball-characterization}, we have
\begin{align*}
    \hball(p,r)
        & ~ = ~ \{ q \in \Omega \ST \distH[\Omega](p,q) \leq r \} \notag\\
        & ~ = ~ \left\{ q \in \Omega \ST \frac{1}{2}\max_{j,k\in[m]}\log \left(\frac{\vctr w_j^\TT \vctr p + c_j}{\vctr w_j^\TT \vctr q + c_j} \frac{\vctr w_k^\TT \vctr q + c_k}{\vctr w_k^\TT \vctr p + c_k}\right) \leq r \right\} \notag \\
       & ~ = ~ \bigcap_{j,k\in[m]} \left\{ q \in \Omega \ST 
       \left(\vctr w_j^\TT \vctr p + c_j\right)\left(\vctr w_k^\TT \vctr q + c_k\right) \leq e^{2r}\left(\vctr w_k^\TT \vctr p + c_k\right)\left(\vctr w_j^\TT \vctr q + c_j\right) \right\}.
\end{align*}

Fixing $p$ and $r$, let $\alpha_{j,k} :=e^{2r}\left(\vctr w_k^\TT \vctr p + c_k\right)$ and $\alpha'_{j,k} := \left(\vctr w_j^\TT \vctr p + c_j\right)$. We have
\begin{align}\label{eq:hilbert-ball-characterization}
    \hball(p,r) 
        ~ = ~ \bigcap_{j,k\in[m]} \left\{ q \in \Omega \ST 0 ~ \leq ~ \left(\alpha_{j,k} \vctr w_j - \alpha'_{j,k} \vctr w_k\right)^\TT \vctr q  + \left(\alpha_{j,k} c_j - \alpha'_{j,k} c_k\right) \right\}.
\end{align}
Let $\vctr w^*_{j,k} = \big(\alpha_{j,k} \vctr w_j - \alpha'_{j,k} \vctr w_k\big)$, and $c^*_{j,k} = \big(\alpha_{j,k} c_j - \alpha'_{j,k} c_k\big)$. Let  $S_{j,k}^H$ be the hyperplane represented by $(\vctr w^*_{j,k},c^*_{j,k})$. Observe that $S_{j,k}^H$ is a linear combination of $L_j$, and $L_k$, and therefore $S^H_{j,k}$ passes through the intersection of $L_j$ and $L_k$ (see Figure~\ref{fig:funk-hilbert-2}(c)). In summary, we have the following:

\begin{lemma}[Hilbert Balls] \label{lem:hball}
Given a set of $m$ hyperplanes, $\mathcal L$ as in Proposition~\ref{prop:birkhoff}, the Hilbert ball $\ballH[\Omega(\mathcal L)](p,r)$ is a polytope bounded by the at most $m(m-1)$ hyperplanes: $S_{j,k}^H$, for $j,k\in[m]$.
\end{lemma}

\section{Hilbert SVM} \label{sec:prob-def}

In this section, we present our solution to the SVM problem in the Hilbert geometry. Recall that our objective is to compute a maximum-margin hyperplane (with respect to the Hilbert distance) that separates two point sets $P^+$ and $P^-$ of total size $n$.

We are given an index set $I = \{1,\ldots,n\}$, and a partition of $I$ into $I^+$, and $I^-$, that is $I^+ \cup I^- = I$, $I^+ \cap I^- = \emptyset$, along with the associated $d$-dimensional point sets: $P^\pm = \{p_i \ST i \in I\}$, $P^+ = \{p_i \ST i \in I^+\}$, $P^- = \{p_i \ST i\in I^-\}$. Throughout this section, we assume $P^+$ and $P^-$ are linearly separable, that is, there is a hyperplane $\hyp K$ such that $P^+ \subset \hyp K^+$, and $P^- \subset \hyp K^-$. (In Section~\ref{sec:soft-margin}, we will consider the general case.)

Define the Hilbert distance of a point $p \in \Omega$ to a hyperplane $\hyp L$ to be $\hilb(\hyp L,p) = \min _{x\in \hyp L} \hilb(x,p)$. We define the \emph{margin} of $P^+$ and $P^-$ with respect to a separating $\hyp L$ as $\min_{p \in P^{\pm}}\hilb (\hyp L, p)$, the minimum distance of any point to $\hyp L$. Finding the maximum margin of any separating hyperplane can be expressed formally as
\begin{equation}
    \text{max} \big\{ \gamma : \exists \hyp K ~\text{such that}~ P^+ \subset \hyp K^+, P^- \subset \hyp K^-,     ~\text{and}~ \hilb(\hyp K,p) \geq \gamma,\; \forall p \in P^{\pm} \big\}.
\end{equation}
Our algorithm will yield an absolute approximation to the optimal solution. In particular, given an \emph{approximation parameter} $\eps > 0$, our algorithm will produce a separating hyperplane that achieves a margin of at least $\gamma - \eps$, where $\gamma$ is the optimal margin.

Assuming $\hyp K$ is represented by the pair $(\vctr w, c)$, this can be expressed equivalently as the following optimization problem:
\begin{align}
\label{opt:init-opt}
\text{maximize: } \quad & \gamma \notag \\
\text{subject to: } \quad 
            & \vctr w^{\TT} \vctr p + c ~ > ~ 0, \quad \forall p \in P^+ \notag \\
            & \vctr w^{\TT} \vctr p +c ~ < ~ 0, \quad \forall p \in P^- \notag \\
            & \hilb(\hyp K,p) ~ \geq ~ \gamma,\quad \forall p \in P^{\pm}, \quad \hyp K \ST (\vctr w, c).
\end{align}

It is not immediately straightforward as to how to solve Opt~\eqref{opt:init-opt}. One difficulty lies in finding a reasonable representation for $\hilb (\hyp K,p)$, the Hilbert distance of a point to a hyperplane. This was done in an earlier work on the SVM in the polytopal Hilbert geometries~\cite{AGV25}, but the size of the representation given there depended on the number of vertices in the Hilbert ball, which grows exponentially with the dimension. Other ways of formulating the SVM problem in hyperbolic spaces usually mirror that in the Euclidean setting, framing it as a non-convex optimization problem, with no theoretical guarantees on the runtime~\cite{cho2019large}. However, by exploiting the geometry of the polytopal Hilbert metric and using the alternative viewpoints of the metrics as presented in Proposition \ref{prop:birkhoff}, we are able to significantly improve upon the prior attempts.


\subsection{Overall Algorithm}

We assume every input parameter is given to us as a rational number, expressed as a fraction of two nonzero $B$-bit integers. We call this representation a \emph{$B$-bit rational number}. Each of the $n$ points $p \in P^{\pm}$ is represented as a $d$-vector $\vctr p$ whose coordinates are $B$-bit rational numbers. The domain $\Omega$ is defined by a set $\mathcal L$ of $m$ hyperplanes, each of whose $(\vctr w, c)$ representation is given as a sequence of $d+1$ $B$-bit rational numbers. We say that the resulting SVM problem instance has \emph{size parameters} $(d, n, m, B)$. The total \emph{bit complexity} of such an instance is $O(d(n+m)B)$.

Before describing the overall algorithm we first establish an upper-bound on $\gamma$ the maximum margin in terms of the bit size of the input. Let $\hyp K$ be the corresponding separating hyperplane for which $\gamma$ is achieved. Observe that for an arbitrary $p_+ \in P^+$, and $p_- \in P^-$, we have  $\hilb (p_-,p_+) \geq \hilb(p_-, \hyp K) + \hilb(p_+, \hyp K) \geq 2\gamma$. Now let the line joining $p_-$ and $p_+$ intersect $\bd \Omega$ in bounding hyperplanes $\hyp L_- : (\vctr w_-,c_-)$, and $\hyp L_+ : (\vctr w_+,c_+)$. Therefore,
\[
    \hilb (p_-,p_+) 
        ~ = ~ \frac{1}{2}\log \frac{d\left(p_+,\hyp L_-\right) d\left(p_-,\hyp L_+\right)}{d\left(p_+,\hyp L_+\right) d\left(p_-,\hyp L_-\right)}
        ~ = ~ \frac{1}{2}\log \frac{\left(\vctr w_-^\TT \, \vctr p_+ + c_-\right)\left(\vctr w_+^\TT \, \vctr p_- + c_+\right)}{\left(\vctr w_-^\TT \, \vctr p_- + c_-\right)\left(\vctr w_+^\TT \, \vctr p_+ + c_+\right)}.
\]

Consider any one of the terms in the numerator, say, $(\vctr w_-^\TT \, \vctr p_+ + c_-)$. Now, $\vctr w_-$, and $p_+$ are given by  $d$ $B$-bit rational numbers, therefore their inner product can have an absolute value of at most $d\cdot2^{2B}$. Hence, the entire term $\vctr w_-^\TT \, \vctr p_+ + c_-$ has an absolute value at most $d\cdot2^{2B} + 2^B$. Recalling that the Hilbert distance is the logarithm of ratios, up to constant factors, $\hilb (p_-,p_+)$ is at most $(8B + 4\log d)\log 2 = O(B+\log d)$, implying that $\gamma \leq M \in O(B+\log d)$. Thus, we have:

\begin{lemma}\label{max-margin}
Given an SVM instance with size parameters $(d, n, m, B)$, the separation margin $\gamma$ for any hyperplane is at most $M(d, n, m, B) = O(B + \log d)$.
\end{lemma}

Our overall algorithm is as follows. Since we are interested in the optimum margin up to an accuracy of $\eps$, we check if $P^+$ and $P^-$ are separable with a margin of at least $r$, where $r$ is an element from the range: $\Gamma = [\eps, 2\eps, \ldots ,M]$, where $M = M(d, n, m, B)$. To check separation for a particular value $r$ we will conduct an (approximate) LP-based feasibility test. We employ this test as part of a parametric binary search over the range $\Gamma$. This implies that we only need to solve $O(\log B + \log \log d + \log(1/\eps))$ many feasibility tests. We next describe this LP-based subroutine.

\subsection{LP-feasibility}

We consider the problem of determining whether it is possible to achieve a given margin $\gamma$, up to an additive error of $\eps$. For a particular margin $r$, recall that $\hball (p,r)$ is the Hilbert ball of radius $r$ centered at $p$. By Lemma~\ref{lem:hball}, $\hball(p,r)$ is a polytope bounded by $m(m-1)$ hyperplanes. (In Section \ref{sec:constr_ball}, we will show how to construct them efficiently.) Let $\mathcal B(r)^+ = \{\hball(p^+,r)\mid p^+ \in P^+\}$, the collection of the Hilbert balls of radius $r$ around the positively labeled points, and define $\mathcal B(r)^-$ analogously for $P^-$. 

The feasibility of $\gamma$ reduces to determining whether the collections of balls $\mathcal B(r)^+$ and $\mathcal B(r)^-$ are linearly separable. For $i\in [n]$, let $\mat A({r,i})$ be an $m(m-1) \times d$ matrix and let $\vctr b(r,i)$ be an $m(m-1) \times 1$ vector such that 
\[
    \vctr x \in \hball(p_i,r) 
        ~ \iff ~ \mat A(r,i)\vctr x + \vctr b(r,i) \geq 0.
\]
We can construct $\mat  A(r,i)$ and $\vctr b(r,i)$ from the hyperplanes $S^H_{j,k} : (\vctr w^*_{j,k},c^*_{j,k})$ described in Lemma~\ref{lem:hball}. In particular, for $t \in [m(m-1)]$, if $(\vctr w^*_t,c^*_t)$ denotes the $t$-th bounding hyperplane of $\hball(p_i,r)$, then $\vctr w^*_t$ is the $t$-th row of $\mat  A(r,i)$ and $c^*_t$ is the $t$-th element of $\vctr b(r,i)$.

Therefore, finding whether $\mathcal B(r)^+$ and $\mathcal B(r)^-$ are linearly separable is equivalent to the following decision problem:
\begin{gather}
    \exists \hyp K \ST (\vctr w, c), \quad \text{such that} \notag\\
        \qquad \forall p_i \in P^+,\notag\\
        \qquad \qquad \forall \vctr x: \mat A(r,i)\vctr x + \vctr b(r,i) \geq 0,\quad \vctr w^{\TT}\vctr x + c \geq 0, \label{eq:primal}\\
        \qquad \forall p_i \in P^-,\notag\\
        \qquad \qquad \forall \vctr x: \mat A(r,i)\vctr x + \vctr b(r,i) \geq 0,\quad \vctr w^{\TT}\vctr x + c \leq 0. \label{eq:primal2}
\end{gather}
For fixed $r$ and $i$, let $\mat A = \mat A(r,i)$ and $\vctr b = \vctr b(r,i)$. Eq.~\eqref{eq:primal} can be rewritten as:
\[
    \min_{\vctr x: \;\mat A\vctr x + \vctr b \geq 0} \vctr w^{\TT}\vctr x + c 
        ~ \geq ~ 0.
\]
By standard LP duality \cite{LPBook}, we have
\[
    \min_{\vctr x: \;\mat A\vctr x + \vctr b \geq 0} \vctr w^{\TT}\vctr x 
    \quad = \quad 
    \max_{\vctr y:\;\mat A^{\TT} \vctr y = \vctr w, \;\vctr y \geq \vctr 0} \vctr -\vctr b^{\TT} \vctr y.
\]
Therefore we can simplify Eq.~\eqref{eq:primal} as follows:
\begin{align*}
\MoveEqLeft
    \forall \vctr x: \mat A\vctr x + \vctr b \geq 0,\quad \vctr w^{\TT}\vctr x + c ~ \geq ~ 0 \\
    & \iff \min_{\vctr x: \;\mat A\vctr x + \vctr b \geq 0} \vctr w^{\TT}\vctr x +c ~ \geq ~ 0\\
    & \iff \max_{\vctr y:\;\mat A^{\TT} \vctr y = \vctr w, \;\vctr y \geq \vctr 0} \vctr -\vctr b^{\TT} \vctr y +c ~ \geq ~ 0 \\
    & \iff \exists \, \vctr y(i):\; \mat A(r,i)^{\TT} \vctr y(i) ~ = ~ \vctr w, 
        ~~ \vctr y(i) \geq \vctr 0, 
        ~~ \vctr b(r,i)^{\TT} \vctr y(i) \leq c.
\end{align*}
Analogously, Eq.~\eqref{eq:primal2} can be simplified to
\[
    \exists \, \vctr y(i):\; \mat A(r,i)^{\TT} \vctr y(i) ~ = ~ \vctr w, 
        ~~ \vctr y(i) \geq \vctr 0, 
        ~~ \vctr b(r,i)^{\TT} \vctr y(i) \geq c.
\]

Hence, to determine whether the collections $\mathcal B(r)^+$ and $\mathcal B(r)^-$ are linearly separable we solve the following LP-feasibility problem in the parameters $\vctr w_{d\times 1}$, $c$, and $\vctr y(i)_{m(m-1) \times 1}$, for $i \in [n]$.
\begin{align}
\text{Minimize:}&&\quad 0 \notag\\
 \text{Subject to:}&\quad   &\vctr b(r,i)^{\TT} \vctr y(i) & ~ \leq ~ c,\quad \forall i \in I^+ \notag\\
    &&\vctr b(r,i)^{\TT} \vctr y(i) & ~ \geq ~ c,\quad \forall i\in I^- \notag\\
    &&\mat A(r,i)^{\TT} \vctr y(i) & ~ = ~ \vctr w, \quad \forall i \in I &\notag\\
    &&\vctr y(i)& ~ \geq ~ \vctr 0,\quad \forall i \in I \notag\\
    &&\vctr 1^{\TT} \vctr w & ~ = ~ 1. \label{opt:dual}
\end{align}
We add the final constraint to avoid the trivial solution in which all the parameters are zero.

\begin{figure}[htbp]
  \centerline{\includegraphics[scale=0.40]{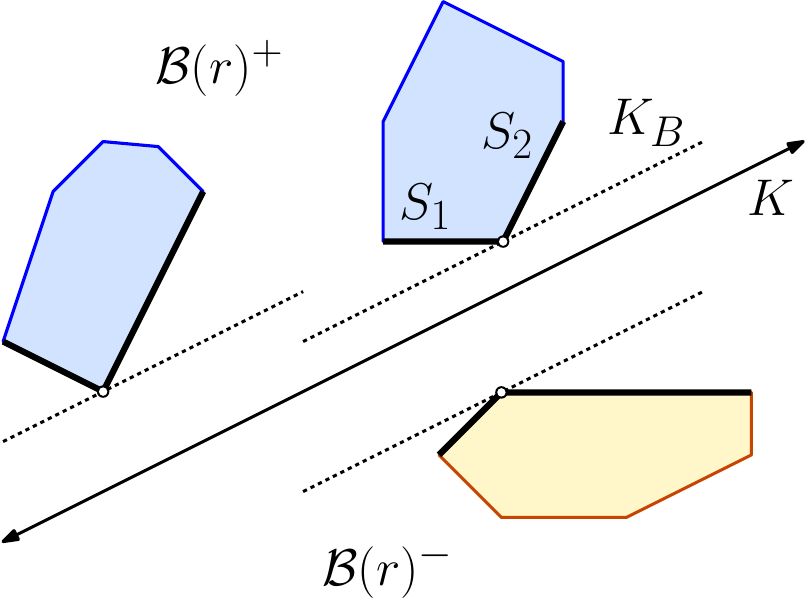}}
  \caption{Intuitive description of the dual: K separates $\mathcal B(r)^+ $ and $ \mathcal B(r)^-$. $K_B$ is a conical combination of $S_1$ and $S_2$ while lying strictly above $K$.}
  \label{fig:intuitive-dual}
\end{figure}

\begin{remark}[Intuitive Description]\label{rem:intuitive LP}
Opt.~\eqref{opt:dual} can be understood intuitively by observing the following. $\mathcal B(r)^+$ and $\mathcal B(r)^-$ being linearly separable implies there exists a hyperplane $\hyp K \ST (\vctr w, c)$, such that for every $B^+ \in \mathcal B(r)^+$, there is a hyperplane $\hyp K_{B^+}$ that is parallel to and above $\hyp K$, and is tangent to some ball in $B^+$. Similarly, for every $B^- \in \mathcal B(r)^-$, there is a hyperplane $\hyp K_{B^-}$ that is parallel to and below $\hyp K$, and is tangent to some ball in $B^-$ (see Figure~\ref{fig:intuitive-dual}).
\end{remark}

A hyperplane $\hyp L$ that is tangent to a polytope $\Omega$, with bounding hyperplanes $\hyp L_1, \ldots, \hyp L_m$, can be written as a positive linear combination (often referred to as the conical combination) of the bounding hyperplanes. 

The vector $\vctr y(i)$ in Opt.~\eqref{opt:dual} corresponds to the factors for this conical combination. The third constraint in Opt.~\eqref{opt:dual} guarantees a hyperplane $\hyp K_B$ parallel to $\hyp K$, along with being a conical combination of the supporting hyperplanes for each  $B \in \mathcal B(r)^+ \cup \mathcal B(r)^-$. The first two constraints ensure $\hyp K_B$ is above or below $\hyp K$, depending on whether $B\in\mathcal B(r)^+$ or $\mathcal B(r)^-$.

\begin{remark}[LP for Funk]\label{LP-funk}
The LP feasibility formulation for the Funk metric is similar to Opt.~\eqref{opt:dual}. We briefly mention the required modifications. By the construction given in Lemma~\ref{lem:fball}, for $i\in [n]$, let $\mat A({r,i})$ be an $m \times d$ matrix and let $\vctr b(r,i)$ be an $m \times 1$ vector such that 
\[
    \vctr x \in \ballF[\Omega](p_i,r) 
        ~ \iff ~ \mat A(r,i)\vctr x + \vctr b(r,i) \geq 0.
\]
For this case, $\vctr y(i)$ will be an $m\times1$ vector. The remainder of the analysis and formulation is exactly the same as in the case of the Hilbert metric.
\end{remark}

\subsection{Constructing Hilbert Balls} \label{sec:constr_ball}

\begin{figure}[htbp]
  \centerline{\includegraphics[scale=0.40]{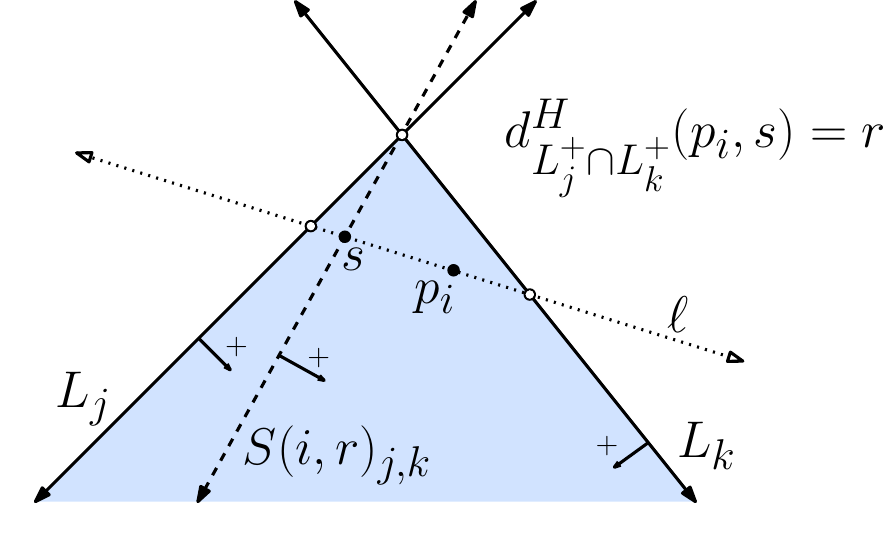}}
  \caption{Constructing $\hball(p_i,r)$.}
  \label{fig:constructing-hball}
\end{figure}

In this section we present a procedure for constructing the bounding hyperplanes for a Hilbert ball, $\hball(p_i,r)$. Recall that $\mathcal L = \{L_i\}_{i=1}^m$ denotes the set of hyperplanes bounding $\Omega$. First, we take two bounding hyperplanes of $\Omega$: $\hyp L_j \ST (\vctr w_j,c_j)$ and $\hyp L_k \ST (\vctr w_k, c_k)$, $j \neq k$, $j,k \in [m]$. For $p_i \in P^{\pm}$, we find $\hyp S(i,r)_{j,k} \ST (\vctr u(i,r)_{(j,k)}, t(i,r)_{j,k})$, the hyperplane at a Hilbert-distance of $r$ from $p_i$ with respect to the boundary defined by $\hyp L_j$ and $\hyp L_k$, such that there exists a line $\ell$ that intersects $\hyp L_j$, $\hyp S(i,r)_{j,k}$, $p_i$, and $\hyp L_k$ in that order (see Figure \ref{fig:constructing-hball}).

It is clear that $\exists~ \alpha: 0\leq\alpha\leq 1$, such that
\begin{equation}\label{eq:ball1}
    \hyp S(i,r)_{j,k} 
        ~ = ~ \alpha \hyp L_j - (1-\alpha) \hyp L_k.
\end{equation}
For any point $s$ on $\hyp S(i,r)_{j,k}$, $s\in \hyp L_j^+ \cap \hyp L_k^+$, it is easy to see that \[
    \frac{d(s,\hyp L_k)}{d(s,\hyp L_j)} 
        ~ = ~ \frac{\alpha}{1-\alpha}.
\]
Solving for $\alpha$ using the following:
\begin{equation}\label{eq:ball2}
    \exp(2\cdot\hilb\left(p_i,\hyp S(i,r)_{j,k}\right)) 
        ~ = ~ \frac{\vctr w_j^{\TT} \vctr p_i + c_j}{\vctr w_k^{\TT} \vctr p_i + c_k}\cdot\frac{\alpha}{1-\alpha} 
        ~ = ~ \exp(2r).
\end{equation}
For $p_i\in \interior(\Omega)$, $0<r<\infty$. Therefore Eq.~\eqref{eq:ball2} can be equivalently written as a linear equation in $\alpha$. Since for $j,k \in [m]$, $j \neq k$, $i\in [n]$, $\hyp S(i,r)_{j,k}$ is a bounding hyperplane for $\hball(p_i,r)$, we can use Eqs.~\eqref{eq:ball1}, and~\eqref{eq:ball2} to construct $\hball(p_i,r)$, for all $p_i \in P^{\pm}$ in $O(ndm^2)$ time.

\begin{remark}[Funk Balls]\label{rem:funk-ball}
We can use a similar approach to construct $B^F_\Omega(p_i,r)$, the Funk ball of radius $r$ around $p_i$. Recall that the Funk ball around $p_i$ is a scaled copy of $\Omega$ around $p_i$. Therefore for a bounding hyperplane of $\Omega$, $\hyp L_j \ST (\vctr w_j, c_j)$, $j \in [m]$, we have $\hyp S(i,r)_j = (\vctr w_j, c^*_j(i,r))$ as a bounding hyperplane of $B^F_\Omega(p_i,r)$ for some constant $c^*_j(i,r)$. We solve for $c^*_j(i,r)$ using the following equation:
\begin{equation}\label{eq:funk-ball-constr}
    \exp(d_\Omega ^F (p_i, \hyp S(i,r)_j)) 
        ~ = ~ \frac{\vctr w_j^{\TT} \vctr p_i + c_j}{\vctr w_j^{\TT} \vctr p_i + c^*_j(i,r)} 
        ~ = ~ \exp(r).
\end{equation}
As before, for any $r \geq 0$, this can be expressed as a linear function in $c^*(i,r)$. Therefore, we can construct $B^F_\Omega(p_i,r)$, for all $p_i\in P^{\pm}$ in $O(nmd)$ time.
\end{remark}

\subsection{Complexity Analysis}

Recall that to obtain $\gamma$ within an additive error of $\eps$, we solve $O(\log B + \log \log d + \log(1/\eps))$ LPs of the form Opt.~\eqref{opt:dual}. We can use any standard interior-point method, such as Khachiyan's Ellipsoid method \cite{khachiyanLP} or Karmarkar's method \cite{KarmarkarLP} to solve Opt.~\eqref{opt:dual} in time that is polynomial in the number of variables, number of inequalities, and bit-length of each variable. We first establish the bit-length required to encode a parameter in our LP-feasibilty problem (Opt.~\eqref{opt:dual}).

Recall that we are searching for $r$ over the range $\Gamma = [\eps,2\eps,\ldots,M]$, with $M = M(d,n,m,B) = O(B + \log d)$. Therefore, the quantity $e^{2r}$ can take on values in the range $\Gamma_{e} = [e^{2\eps}, e^{4\eps},\ldots,e^{2M}]$. Instead of fixing $r$'s value exactly over $\Gamma$, it is sufficient for our purposes if we search for $r$ over $[r_1, r_2, \ldots, r_{M/\eps}]$, where $(i-1) \eps \leq r_i < i \eps$.

The minimum difference between two elements in $\Gamma_e$ is at least $e^{4\eps} - e^{2\eps} > \eps$. (This can be verified using Taylor's series for $e^x$, at $x=0$.) Therefore, to represent $\Gamma_e$, it is sufficient for our purposes to find a bit-representation which allows for a maximum value $e^{2M}$ up to a resolution of $\eps$. This can be done by using $O(M)$ bits for the integer part, and $O(\log (1/\eps))$ bits for the fractional part. Hence, it is sufficient to represent the elements of $\Gamma_e$ as $O(M+\log (1/\eps)) = O(B + \log d + \log(1/\eps))$-bit rational numbers.

Using Eq.~\eqref{eq:ball2}, we solve for and represent $\alpha$, using $O(B + \log d + \log(1/\eps))$-bit rational numbers. Therefore, using Eq.~\eqref{eq:ball1} every parameter of $\hyp S(i,r)_{j,k}$, which acts as a bounding hyperplane for $\hball (p_i, r)$, can also be represented using $O(B + \log d + \log(1/\eps))$-bit rational numbers.

Now, the number of variables in Opt.~\eqref{opt:dual} is $O((n+d)m^2)$, with the number of inequalities being $O(n(m^2+d))$. Hence, we can determine if Opt.~\eqref{opt:dual} is feasible in $O(\text{poly}(n,m,d)(B + \log d + \log(1/\eps))$ time, for a fixed $r$. And since we solve $O(\log B + \log \log d + \log(1/\eps))$ such LPs, our overall running time is $O(\text{poly}(n,m,d,B,\log(1/\eps))$. This proves our main result in Theorem \ref{thm:main-result}. Recall that the input bit-length is $d(n+m)B$ implying the overall running time is also polynomial in the input bit-length and the log of the desired accuracy.

\section{Extensions}\label{sec:extensions}

In this section, we will explore two extensions to binary classification and SVM problem in the Hilbert geometry. The first is a soft-margin classifier, which is applicable when the point sets are not linearly separable, and the second is a binary classifier based on nearest neighbors.

\subsection{A Soft-Margin Classifier} \label{sec:soft-margin}

Up to this point we assumed $P^+$, and $P^-$ are linearly separable. If they are not, a common approach is to consider a hyperplane classifier with appropriate penalties associated with misclassification of points in the training set. In this section we propose a formulation for such a ``soft'' classifier.

We modify our LP-feasibility problem: Opt~\eqref{opt:dual} to introduce penalties in the form of slack variables. Reiterating the observation from Remark \ref{rem:intuitive LP}, we note that the first two constraints of Opt~\eqref{opt:dual} enforce the condition that, for $p\in P^+$, $\hball(p,r)$ is completely above the separator $\hyp K \ST (\vctr w, c)$. Instead, we only require that $\hball (p,r)$ is above $\hyp K' \ST (\vctr w, c+\xi)$ for some positive $\xi$. Here $\xi$ works as a penalty for the improper separation of $\hball(p,r)$. We similarly set individual penalties for each point. 

Next, we want the following desirable property: points that are closer to the boundary of $\Omega$ should pay a heavier penalty for being mis-classified. It is because for a fixed point $q$, $\hilb (p_i,q)$ scales inversely with $p_i$'s distance to the boundary. Therefore we scale the penalty associated with $p_i$ by $\omega_i$, defined as the inverse of its distance to the boundary: 
\[
\omega_i = \frac{1}{\min _{j \in [m]}\hilb (L_j,p_i)}
\]

Finally we propose the following LP in the parameters $\vctr w$, $c$, and $\xi_i$, $\vctr y(i)$, for $i \in [n]$, to minimize the overall weighted penalty for a fixed Hilbert radius $r$:
\begin{align}
\text{Minimize:}&\quad &\Xi_r & ~ = ~ \sum_{i\in [n]}\omega_i \xi_i \notag\\
 \text{Subject to:}&\quad   &\vctr b(r,i)^{\TT} \vctr y(i) & ~ \leq ~ c+\xi_i,\quad \forall i \in I^+ \notag\\
    &&\vctr b(r,i)^{\TT} \vctr y(i) & ~ \geq ~ c-\xi_i,\quad \forall i \in I^- \notag\\
    &&\mat A(r,i)^{\TT} \vctr y(i) & ~ = ~ \vctr w, \quad \forall i \in I &\notag\\
    &&\vctr y(i) & ~ \geq ~ \vctr 0,\quad \forall i \in I \notag\\
    &&\vctr 1^{\TT} \vctr w & ~ = ~ 1 \notag\\
    &&\xi_i & ~ \geq ~ 0,\quad \forall i \in I. \label{opt:SVC}
\end{align}

Let the optimum value derived from Opt.~\eqref{opt:SVC} be $\Xi_r$, and the corresponding separating hyperplane; $\hyp K_r =(\vctr w,c)$. Recall that $r$ attains values over $\Gamma_r = [r_1, r_2, \ldots, r_{M/\eps}]$, where $r_i$'s are any values satisfying $(i-1) \eps \leq r_i < i \eps$, and $M = O(B + \log d)$. To find an appropriate classifier $\hyp K^*$, we take 
\[
    \hyp K^* ~ = ~ \hyp K_s, \qquad\text{where}\qquad 
    s ~ = ~ {\arg\max}_{i\in [M/\eps]} \left(r_i - C~\frac{\Xi_{r_i}}{n}\right).
\]
Note that $C$ is a user defined-constant, a weight associated with improper separation. A higher value would be partial towards proper separation, while a lower value towards a larger margin.

A slight drawback of this approach is that it no longer suffices to perform a binary search on the range $\Gamma_r$. Rather, we need to perform a grid search, i.e. solve Opt.~\eqref{opt:SVC} for every value of $r\in\Gamma_r$ in the worst case. This implies the overall running time for the soft-margin classifier is $O(\text{poly}(n,m,d,B,1/\eps))$.

\subsection{A Nearest Neighbor-based Classifier}\label{sec:nn-class}

In this section, we consider an alternative approach to binary classification in the Hilbert geometry. Rather than using a separating hyperplane, we select two sites $q^+$ and $q^-$, and a point $p$ is classified according to which of these two sites it is closer to. In the Euclidean setting, the bisector between two points is a hyperplane, and this is no different from SVM. However, in Hilbert, bisectors are generally not hyperplanes. (An analysis of the structure of Hilbert bisectors in 2-dimensional case was presented in \cite{gezalyan2023voronoi}, where it was shown that they are piecewise conics.)

As before, we assume that Hilbert geometry is defined with respect to a polytope $\Omega$ in $\RE^d$ defined by $m$ bounding hyperplanes $\mathcal L = \{L_i\}_{i=1}^m$. Extending results of de la Harpe~\cite{Har93}, Vernicos presented an elegant isometry from the polytopal Hilbert geometry $(\Omega,\hilb)$ to a finite dimensional normed vector space~\cite{vernicos2014hilbert}. In particular, he presented an isometric mapping $f_{\Omega} : (\Omega, \hilb) \rightarrow (\RE^{m-1},||\cdot||_{\Sigma_{2m}})$, where $||\cdot||_{\Sigma_{2m}}$ is a norm defined by a regular polytope $\Sigma$ bounded by $2m$ facets in $\RE^{m-1}$. In other words, the embedding space is $\RE^{m-1}$ with $\Sigma_{2m}$ as the unit ball.

Although one might be tempted to solve the SVM problem in the embedding space, the result would be far from ideal. Since $f_{\Omega}$ is nonlinear, a linear classifier $\widehat{\hyp K}$ in the embedding space would not translate to a simple classifier in the input space. Moreover, since $f_{\Omega}$ is not a bijection it is not clear how complex the intersection of $\widehat{\hyp K}$ with the image of $\Omega$, $f_{\Omega}(\Omega)$, would be.

Instead, we propose a simple nearest neighbor-based classifier in the embedding space. Given $\Omega$, let $q_i$, $Q^+$, and $Q^-$ denote the images of $p_i$, $P^+$, and $P^-$ under the embedding $f_{\Omega}$. ($q_i$ for all $i\in[n]$ can be computed in $O(m^2)$ time by solving $m$ linear equations) We find two representative centers $c^+$, and $c^-$, for $Q^+$, and $Q^-$ respectively such that the following is maximized:
\begin{equation}
    \beta 
        ~ = ~ \max_{c^+,c^-\in \RE^{m-1}} \left(
            \min_{q^+ \in Q^+}
                \frac{\min_{q^- \in Q^-}\norm{\vctr q^+ - \vctr q^-}_\Sigma}{\norm{\vctr q^+ - \vctr c^+}_\Sigma},
            \min_{q^- \in Q^-}
                \frac{\min_{q^+ \in Q^+}\norm{\vctr q^- - \vctr q^+}_\Sigma}{\norm{\vctr q^- - \vctr c^-}_\Sigma}
        \right) 
\end{equation}
Stated differently, $\beta$ is the value such that any point $q$ is at least $\beta$ times away from its nearest neighbor in the opposite class, as from its representative. This implies that the points can be correctly classified using an approximate nearest-neighbor algorithm that achieves an approximation factor of at most $\beta$. 

This can be rewritten as an optimization problem over the variables $\vctr c^+$, $\vctr c^-$, and $1/\beta$
\begin{align}\label{opt:nnclass}
    \text{Minimize:}&\quad &\frac{1}{\beta}& \notag\\
    \text{Subject to:}
         &\quad &\norm{\vctr q^+ -\vctr c^+}_\Sigma ~ \leq ~ \frac{1}{\beta}\norm{\vctr q^+ -\vctr q^-}_\Sigma, & \quad \forall q^+\in Q^+, \forall q^- \in Q^- \notag\\
         &\quad &\norm{\vctr q^- -\vctr c^-}_\Sigma ~ \leq ~ \frac{1}{\beta}\norm{\vctr q^- -\vctr q^+}_\Sigma, & \quad \forall q^+\in Q^+, \forall q^- \in Q^-
\end{align}

We note that $\norm{\vctr q^+ - \vctr q^- }_\Sigma$ in the above optimization is a pre-determined constant. If we let $W = \{\vctr w_1,\vctr w_2, \ldots,\vctr w_{2m}\}$ denote the directions determining the norm polytope $\Sigma$, then the condition that $\norm{\vctr x}_\Sigma \leq t$ can be expressed as a set of linear inequalities: $\vctr w_i^{\TT} \vctr x \leq t$, for $i \in [2m]$. Therefore, Opt.~\eqref{opt:nnclass} is an LP in $O(m)$ variables, and $O(nm)$ inequalities.

\bibliography{shortcuts,hilbert}

\end{document}